\documentclass[aps,prr,reprint,superscriptaddress,floatfix]{revtex4-2}

\usepackage{blindtext}
\usepackage{amssymb}
\usepackage{amsmath}
\usepackage{mathtools}
\usepackage{cases}
\usepackage[normalem]{ulem}
\usepackage{physics}
\usepackage{xcolor}
\usepackage{calc}
\usepackage{hyperref}
\hypersetup{colorlinks, allcolors=blue}
\usepackage[capitalise]{cleveref}
\usepackage{tikz}
\usepackage{comment}
\definecolor{mycolor}{rgb}{0.1, 0.1, 0.7} 
\hypersetup{
    plainpages=true,
    breaklinks=true,
    colorlinks=true,
    linkcolor={mycolor},
    citecolor={mycolor},
    urlcolor={mycolor},
}
\usetikzlibrary{
    math,
    arrows,
    arrows.meta,
    decorations.pathmorphing, 
    decorations.shapes,
    positioning,
    calc,
    shapes,
    shapes.geometric
}

\usepackage{appendix}
\usepackage{physics}

\makeatletter
\def\maketitle{
\@author@finish
\title@column\titleblock@produce
\suppressfloats[t]}
\makeatother

\makeatletter
\AddToHook{cmd/appendix/before}{\def\cref@section@alias{appendix}\def\cref@subsection@alias{appendix}}
\makeatother

\begin{document}

\title{Programmable generation of flying cat-qubits}

\author{Cecilia Erneman}
\author{Zeidan Zeidan}
\author{Göran Johansson}
\author{Maryam Khanahmadi}
\email{m.khanahmadi@chalmers.se}
\affiliation{Department of Microtechnology and Nanoscience (MC2),
Chalmers University of Technology, 41296 Gothenburg, Sweden}

\date{\today}

\begin{abstract}

We propose a framework for the direct generation of flying cat-qubit states from vacuum using time-dependent two-photon drives in nonlinear bosonic systems. We study both Kerr-based and two-photon-dissipation-based generation. By engineering Kerr nonlinearity, two-photon driving, and dissipation, we demonstrate logical control of a cat qubit during its generation and emission, while its quantum information is simultaneously shared between the nonlinear system and the propagating output field. We further analyze the effects of photon loss and pure dephasing, showing that both the state generation and logical control remain robust under realistic noise conditions. These results provide a route toward programmable bosonic quantum networks and future propagating error-correctable encodings.
\end{abstract}

\maketitle

\section{Introduction}

Itinerant quantum states are a key resource for quantum networking and distributed quantum computing, where quantum information must be transmitted between distant nodes with high fidelity. Among the various encoding strategies, bosonic encodings based on Schrödinger cat states have attracted considerable attention because of their resilience to noise \cite{terhal2020towards,PhysRevLett.119.030502,joshi2021quantum}. Superconducting circuits, owing to the strong nonlinearities provided by Josephson junctions \cite{devoret2004implementing}, introduce high degrees of control for non-linear interactions to generate and control such non-Gaussian states \cite{ding2025quantum,lescanne2020exponential,axline2018demand,pfaff2017controlled,khanahmadi2025environment,PhysRevResearch.5.043071,eriksson2024universal,puri2020bias,mirrahimi2016cat,reglade2024quantum,Gautier_2022,he2023fast,PhysRevX.14.041049,PhysRevLett.111.120501}.

Bosonic cat states are useful for enabling fault-tolerant quantum computing \cite{PhysRevX.9.041053,putterman2025hardware,PRXQuantum.3.010329}, quantum metrology \cite{PhysRevA.107.013705,72f2-tgwp}, and quantum networking tasks such as quantum state transfer, entanglement distribution, and modular quantum computing \cite{axline2018demand,khanahmadi2025environment,PhysRevResearch.5.043071,teoh2025robust,PhysRevA.99.023838}.

In this work, we focus on generating high-fidelity propagating cat states for networking between distant quantum nodes. Conventionally, these states are first prepared and stored in a cavity before being released into a transmission line through a tunable output coupler. More recently, theoretical proposals have shown that cat states can instead be generated and emitted simultaneously, eliminating the need for a separate release stage \cite{PhysRevA.99.023838,khanahmadi2025environment}.

While existing protocols can generate propagating cat states with high fidelity, they generally provide limited control over the logical state encoded in the emitted field. High-fidelity quantum networking requires quantum nodes that provide flexible control over logical bosonic states, enabling protocols such as deterministic entanglement generation, quantum state transfer, and distributed quantum computation. In the cat-state encoding, this corresponds to controlling both the amplitudes and the relative phase of the logical basis states. Developing such a programmable bosonic source is an important step toward realizing propagating cat qubits for quantum networking applications.

In this work, we propose a protocol for the direct generation and emission of flying cat qubits with arbitrary logical-state control. We investigate both Kerr-based and dissipation-based generation schemes and demonstrate logical X and Z operations on the emitted cat qubit. Finally, we analyze the robustness of the protocol against photon loss and pure dephasing. 

The paper is organized as follows. Section~\ref{cat-state} introduces the cat-state encoding and the generation protocols. Section~\ref{Output} describes the characterization of the emitted temporal mode and the reconstruction of the propagating quantum state. Section~\ref{Counter} presents the shortcut-to-adiabaticity approach for accelerating the generation process. Section~\ref{results} presents the numerical results, including the effects of photon loss and dephasing, and Section~\ref{summary} concludes the paper.


\section{Flying Cat Qubits and Generation Protocols}\label{cat-state}
\subsection{Two-component cat qubit}
\begin{figure}[t!]
    \centering
    \includegraphics[width=\linewidth]{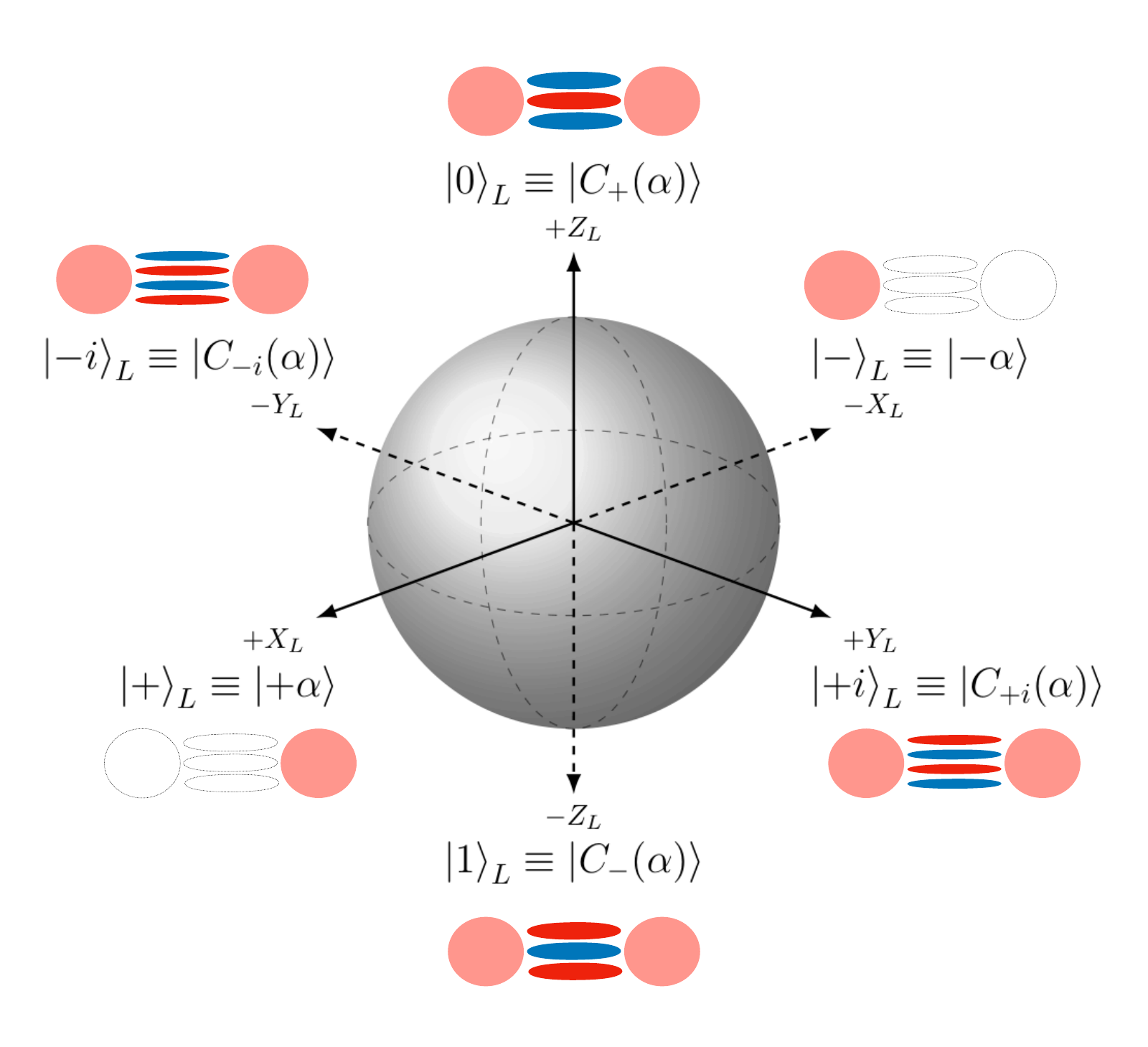}
    \caption{Bloch-sphere representation of the two-component cat logical qubit. The logical computational basis is chosen as $\ket{0_L}\equiv\ket{C_+(\alpha)}$ and $\ket{1_L}\equiv\ket{C_-(\alpha)}$, defining the $\pm Z_L$ axis. The remaining cardinal states correspond to eigenstates of the logical Pauli operators: $\ket{\pm}_L \equiv \ket{\pm \alpha}$ and $\ket{\pm i}_L \equiv \ket{C_{\pm i}(\alpha)} \propto \ket{\alpha}\pm i\ket{-\alpha}$, along $\pm X_L$ and $\pm Y_L$, respectively.}
    \label{fig:cat-bloch-sphere}
\end{figure}

The two-component cat states \cite{mirrahimi2014dynamically, Puri2017NPJQI} are a bosonic encoding built from a superposition of coherent states 
\begin{equation}\label{eq:2-component-cat-states-def}
        \ket{C_{\pm}(\alpha)} = \frac{\ket{\alpha} \pm \ket{-\alpha}}{\mathcal{N}_\pm},
\end{equation}
with coherent state amplitude $\alpha$ and normalization constant $\mathcal{N}_\pm = \sqrt{2(1 \pm e^{-2|\alpha|^2})}$. The even- and odd-parity cat states define the logical basis of the two-component cat qubit, with $\ket{0_L}\equiv\ket{C_+(\alpha)}$ and $\ket{1_L}\equiv\ket{C_-(\alpha)}$. 
\cref{fig:cat-bloch-sphere} shows the schematic of the cat-qubit Bloch sphere \cite{mirrahimi2014dynamically}, where an arbitrary superposition of a cat qubit can be written as
\begin{equation}\label{eq:logical-2-comp-cat-qubit-def}
    \ket{\psi_L(\alpha, \theta, \varphi)} = \cos(\theta/2) \ket{C_+(\alpha)} + e^{i\varphi}\sin(\theta/2)  \ket{C_-(\alpha)},
\end{equation}
with polar angle $\theta \in [0, \pi]$ and azimuthal angle $\varphi \in [0, 2 \pi)$. Henceforth, we denote the probabilities of the even and odd logical basis states by $\{ p_+ = \cos^2(\theta/2),p_- = \sin^2(\theta/2)\}$. 

The large separation between the coherent states $\ket{\alpha}$ and $\ket{-\alpha}$ leads to an exponentially small overlap,
$\bra{\alpha}\ket{-\alpha}=e^{-2|\alpha|^2}$,
which suppresses unwanted transitions between the coherent components. Equivalently, in the parity-cat basis, this corresponds to an exponential suppression of logical phase-flip errors \cite{ grimm2020stabilization,lescanne2020exponential,puri2020bias,Gautier_2022}. Although the two-component cat state is not fully error-correctable against photon loss, it provides the simplest bosonic logical basis for studying programmable flying qubits. In this encoding, one can directly investigate the generation of arbitrary logical states, parity control, leakage from the cat manifold, and the single-modeness of the emitted field. Therefore, controlling two-component flying cat qubits is an important step toward more complex propagating bosonic encodings.

\subsection{Generation Protocol}
\label{subsec:generation_protocols}
\begin{figure}[t!]
    \centering
    \includegraphics[width=\linewidth]{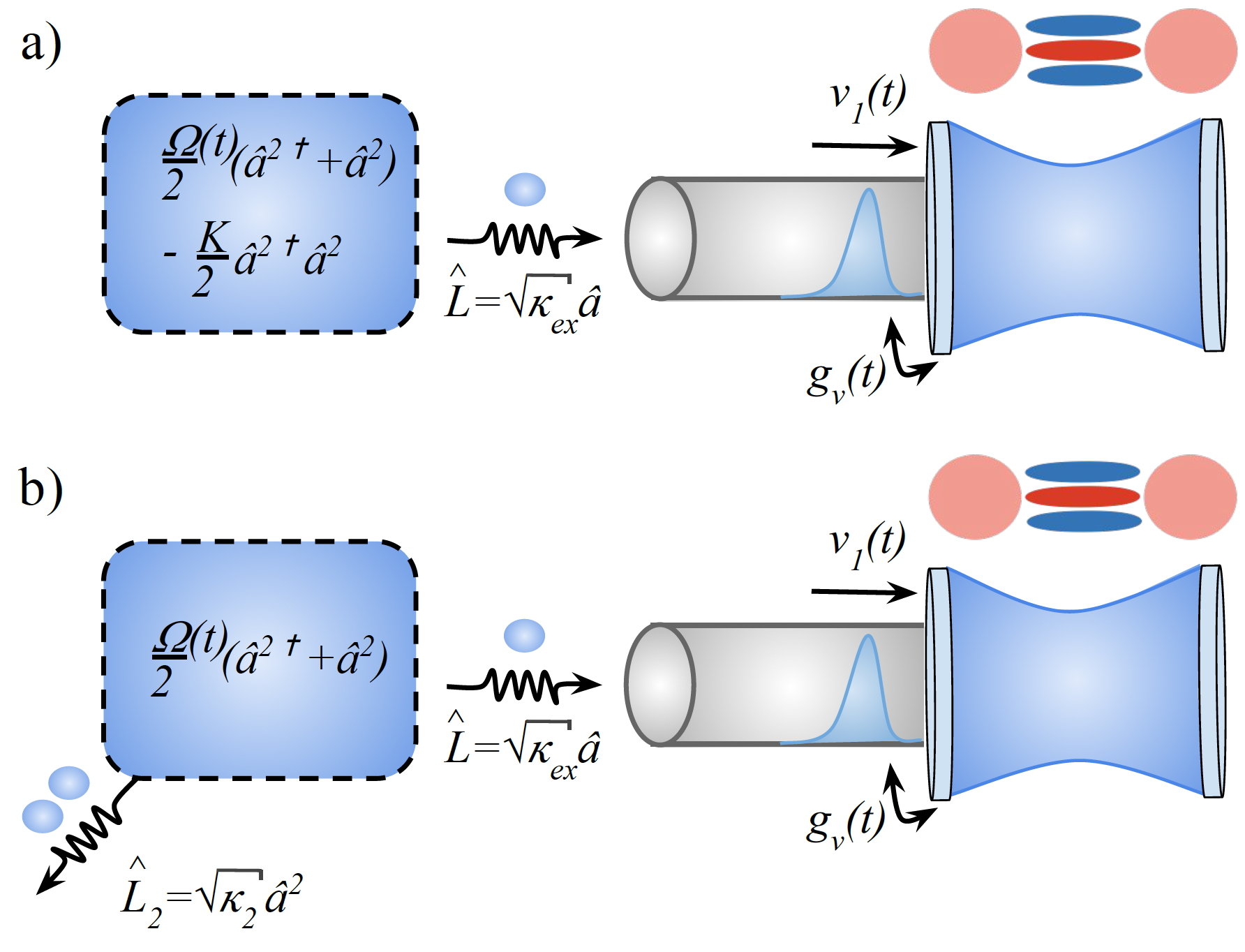}
    \caption{Schematic of the flying cat-qubit generation protocols based on either a Kerr nonlinear Hamiltonian (a) or engineered two-photon dissipation (b). The source is strongly coupled to a waveguide and driven by a time-dependent two-photon pump, $\Omega(t)(\hat{a}^{\dagger 2}+\hat{a}^2)$, which generates and controls the logical state of the emitted cat qubit. The resulting wavepacket occupies the dominant temporal mode $v_1(t)$. A linear catching cavity with time-dependent coupling $g_v(t)$ is used to capture this mode and reconstruct the quantum state of the propagating cat qubit.}
    \label{fig:setup}
\end{figure}

To generate a flying cat qubit, we consider the quantum source system  as a non-linear quantum mode described with ladder operators $\{\hat{a},\hat{a}^\dagger\}$, strongly coupled to a one-dimensional waveguide through the output channel $\hat{L} = \sqrt{\kappa_\mathrm{ex}} \hat{a}$ with coupling decay rate $\kappa_\mathrm{ex}$ and controlled by the time-dependent two-photon drive $\Omega(t)$, to generate quantum state \cref{eq:logical-2-comp-cat-qubit-def} into propagating mode; see  \cref{fig:setup}. 

In this work, we consider two generation schemes: Hamiltonian-based generation using a Kerr-nonlinear parametric oscillator, and dissipation-based generation using engineered two-photon loss. In both cases, the temporal profile of the emitted field is controlled through the two-photon drive envelope $\Omega(t)$; the performance of shaping is investigated in \cref{subsec:shaping-results}.


\textit{Hamiltonian-based generation---}A cat qubit can be generated using a Kerr parametric oscillator with Hamiltonian
\begin{equation}\label{eq:H-KPO-Nakamura-def}
\hat{H}_\mathrm{KPO}(t) = \frac{\Omega(t)}{2}(\hat{a}^{\dagger 2} + \hat{a}^2) - \frac{K}{2}\hat{a}^{\dagger 2} \hat{a}^2,
\end{equation}
where $\Omega(t)$ is the two-photon drive amplitude and $K$ is the Kerr nonlinearity. The low-energy manifold of \cref{eq:H-KPO-Nakamura-def} is spanned by cat states $\ket{C_{\pm}(\alpha(t))}$, with coherent-state amplitude $\alpha(t)=\sqrt{\Omega(t)/K}$. This Hamiltonian can be implemented in superconducting circuits using a $\lambda/4$ resonator terminated by a flux-tunable transmon. By applying a parametric flux drive at approximately twice the resonator frequency, the two-photon interaction in \cref{eq:H-KPO-Nakamura-def} can be engineered \cite{Puri2017NPJQI,Wang_2019,Krantz_2019,PhysRevA.107.042407,PhysRevX.14.041049}.

\textit{Dissipation-based generation---} 
An alternative realization is provided by a driven-dissipative setup, where a linear resonator is coupled to a lossy buffer mode, typically through three-wave mixing interaction, to engineer two-photon dissipation \cite{mirrahimi2014dynamically, Gautier_2022,PRXQuantum.4.040316,reglade2024quantum}.
In this approach, the quantum system is subjected to two-photon dissipation $\hat{L}_{2} = \sqrt{\kappa_2}\hat{a}^{2}$, where including two photon-drive $\hat{H}_d(t)= \Omega(t)/2 (\hat{a}^{\dagger 2} + \hat{a}^2)$, the non-Hermition Hamiltonian $\hat{H}_\mathrm{nh} = \hat{H}_d(t) - \frac{i}{2} \hat{L}_{2}^\dagger \hat{L}_{2}$
\cite{castin2008wavefunctionapproachdissipative,terhal2020towards}, generates two-component cat state.

\textit{Release and propagate the cat qubit---} 
As mentioned before, with drive profile $\Omega(t)$, we inject the photon into the quantum system, and due to strong dissipation $\kappa_{\text{ext}}$, the cavity field simultaneously leaks to the waveguide. The Lindblad master equation of Hamiltonian-based generation is described by
\begin{align}\label{ev1}
    \dot{\hat{\rho}}_s = -i[\hat{H}_\mathrm{KPO}(t),\hat{\rho}_s] + \kappa_{\text{ex}}\big( \hat a \hat{\rho}_s \hat a^\dagger-\tfrac{1}{2}\{\hat{\rho}_s,\hat a^\dagger \hat a\}\big),
\end{align}
and by employing two-photon dissipation, the system evolves under the process 
 \begin{align}\label{ev2}
        \dot{\hat{\rho}}_s = -i[\hat{H}_d(t),\hat{\rho}_s] &+ \kappa_{\text{ex}}\big( \hat a\hat{\rho}_s \hat a^\dagger-\tfrac{1}{2}\{\hat{\rho}_s,\hat a^\dagger \hat a\}\big) \nonumber\\
        &+\kappa_{2}\big(2 \hat a^2 \hat{\rho}_s \hat a^{\dagger 2}-\{\hat{\rho}_s,\hat a^{\dagger2} \hat a^2\}\big).
 \end{align}


\subsection{Characteristics of propagating mode}\label{Output}

To characterize the emitted flying cat qubit and generation fidelity, we utilize the mode decomposition method to find the dominant temporal mode. Using the quantum regression theorem \cite{PhysRevLett.123.123604,PhysRevA.102.023717}, we can evaluate the first-order correlation function of the output field,  $g^{(1)}(t,t') = \langle \hat L(t)^\dagger \hat L(t')\rangle=\kappa_\mathrm{ex} \langle \hat{a}^\dagger(t)\hat{a}(t')\rangle$. The temporal-mode structure of the output field is determined by the eigenmode decomposition $g^{(1)}(t,t') = \sum_i n_i v_i^*(t) v_i(t')$, where $\{v_i(t)\}_i$ is an orthonormal set of temporal modes and $n_i$ is the mean photon number in mode $v_i$ \cite{PhysRevA.102.023717}. We consider the dominant mode $v_1(t)$ to have the largest photon occupation number $n_1$.

To reconstruct the quantum state carried by the dominant temporal mode $v_1$, we introduce a catching cavity with ladder operators $\{\hat{a}_v,\hat{a}_v^\dagger\}$, coupled to the source through a cascaded interaction \cite{PhysRevLett.123.123604,PhysRevA.102.023717}, as shown in \cref{fig:setup}. By controlling the coupling strength
\begin{equation}\label{eq:gv-def}
g_{v}(t) = - \frac{v_1^*(t)}{\sqrt{\int_0^t dt' |v_1(t')|^2}},
\end{equation}
the catching cavity, described by the Lindblad operator $\hat L_v(t)=g_v^*(t)\hat a_v$, can capture the temporal mode $v_1$. Consequently, the final state of the catching cavity corresponds to the quantum state carried by that propagating mode \cite{PhysRevLett.123.123604}.

Utilizing the SLH formalism \cite{combes2017slh} for cascaded systems, the total dynamics of the source and catching cavity are described by
\begin{equation}\label{eq:ME-rho_sv}
    \Dot{\hat{\rho}}_{\mathrm{sv}}(t) = -i[\hat{H}_{\mathrm{sv}}(t), \hat{\rho}_{\mathrm{sv}}(t)] + \mathcal{D}(\hat{L}_{\mathrm{sv}}(t)) \hat{\rho}_{\mathrm{sv}}(t),
\end{equation}
with cascaded Hamiltonian \cite{PhysRevA.102.023717}
\begin{equation}\label{eq:cascaded-Hamiltonian}
    \begin{split}
        \hat{H}_{\mathrm{sv}}(t) & = \hat{H}_{\mathrm{s}}(t) + \hat{H}_{\mathrm{cross}}(t) \\
        & = \hat{H}_{\mathrm{s}}(t) + \frac{i \sqrt{\kappa_{\mathrm{ex}}}}{2} (g_{v}^*(t) \hat{a}^\dagger \hat{a}_{v} - g_{v}(t)\hat{a}_{v}^\dagger \hat{a}),
    \end{split}
\end{equation}
and the total Lindblad operator
\begin{equation}\label{eq:cascaded-jump-operator}
    \begin{split}
        \hat{L}_{\mathrm{sv}}(t) & = \hat{L} + \hat{L}_v(t) \\
        & = \sqrt{\kappa_{ex}} \hat{a} + g_{v}^*(t) \hat{a}_{v}.
    \end{split}
\end{equation}

\subsection{Logical rotation of flying qubit}\label{subsec:Logical-rot-of-flying-cat-qb}

In both generation scenarios, the quantum sources start from vacuum and evolve under the two-photon drive $\Omega(t)$. Due to parity preservation of the evolution, \textit{i.e.}, the two-photon pump, Kerr nonlinearity, and two-photon dissipation all change the photon number by two, thus the generated cat state has even parity. However, to encode quantum information and investigate the possibility of implementing quantum gates on propagating cat states, an arbitrary superposition of the logical states is required. Therefore, parity must be broken. This can be achieved by applying a single-photon drive, which acts as a logical rotation gate on the cat Bloch sphere \cite{PhysRevA.99.023838,mirrahimi2014dynamically,PhysRevResearch.6.013192,PhysRevA.94.033841}.
We thus introduce a time-dependent and complex single-photon drive,
\begin{equation}\label{eq:H-eps-w-phieps}
    \hat{H}_\varepsilon(t) = \varepsilon(t)(e^{i\phi_\varepsilon}\hat{a} + e^{-i\phi_\varepsilon}\hat{a}^\dagger),
\end{equation}
where $\varepsilon(t)$ and $\phi_\varepsilon$ correspond to the drive envelope and the drive phase, respectively. 

For $\phi_\varepsilon=0$, this reduces to a real quadrature drive proportional to $\propto \hat{a}+\hat{a}^\dagger$, which couples the even and odd cat states and implements the logical $X$-axis rotation studied in \cref{subsec:X-rot}. To establish arbitrary rotation within the logical subspace, control over the relative phase $\varphi$ in \cref{eq:logical-2-comp-cat-qubit-def} is required. In the stationary mode generation, a small detuning can provide such a logical rotation. However, for propagating cat states, detuning leads to a multimode output field and is therefore undesirable.

Instead, we consider a single-photon drive with phase $\phi_\epsilon\neq 0$. In this case, the logical rotation is the combination of $X$ and $Y$ rotations that provides an effective $Z$ rotation, allowing control of the relative phase $\varphi$; the simulation results are provided in \cref{results}. It is important to note that, to ensure that the dynamics remain within the cat manifold and to suppress leakage to higher excited states, the single-photon drive must remain sufficiently weak; i.e., $\epsilon(t)\ll 4K|\alpha|^2$ \cite{mirrahimi2014dynamically,grimm2020stabilization}.

We need to mention that an alternative approach for the arbitrary superposition would be to initialize the source oscillator in a superposition $\propto(p_0 \ket{0}+ p_1 e^{i\varphi}\ket{1})$ and apply the evolution equations \cref{ev1} or \cref{ev2} to have the arbitrary superposition. However, we do not pursue this approach here. In our simulations, these initial superposition protocols produced more multimode output fields. In addition, such an initialization requires a longer overall sequence, since the initial superposition must be prepared before the cat-generation ramp. Furthermore, because the source is strongly coupled to the waveguide, the initial state partially decays during the preparation process and does not lead to the desired cat-state superposition determined by the amplitudes $p_{\pm}$.

We therefore focus on the single-photon-drive protocol, where the logical rotation is implemented during the generation and release process. This approach provides full arbitrary control within the logical subspace and demonstrates the possibility of applying quantum gates directly to propagating cat states while only part of the state is inside the cavity.

\subsection{Shortcut to adiabaticity}\label{Counter}
The time-dependent two-photon pump and single-photon drive inevitably cause non-adiabatic transitions out of the cat manifold, degrading the fidelity of the emitted flying cat qubit. The shortcut to adiabaticity (STA) technique suppresses these transitions by adding a counteradiabatic term to the Hamiltonian, which enforces adiabatic evolution without slowing the protocol \cite{PhysRevLett.111.100502,PhysRevResearch.6.013192}. In general, the counteradiabatic Hamiltonian takes the form \cite{PhysRevLett.111.100502}
\begin{equation}\label{eq:H-STA}
    \hat{H}_\mathrm{STA} = \frac{i}{2} \sum_{n}\left(\ket*{\dot{\phi}_n}\bra{\phi_n} - \ket{\phi_n}\bra*{\dot{\phi}_n}\right),
\end{equation}
where $\ket{\phi_n}$ is the $n$th eigenstate of the Hamiltonian of interest.

Applying \cref{eq:H-STA} to the cat states of the KPO in \cref{eq:H-KPO-Nakamura-def}, whose eigenstates are $\ket{C_\pm(\alpha)}$ with $\alpha(t) = \sqrt{\Omega(t)/K}$, yields the counteradiabatic Hamiltonian \cite{PhysRevA.99.023838}
\begin{equation}\label{eq:H-counter-Nakamura}
    \hat{H}_\mathrm{STA} = \frac{i\dot{\Omega}(t)}{2\Omega(t)}\tanh\left(\frac{\Omega(t)}{K}\right)\left(\hat{a}^{\dagger 2}-\hat{a}^2\right),
\end{equation}
which is proportional to the rate of change of the two-photon pump $\dot{\Omega}$.

In addition to the KPO Hamiltonian, we need to find the counter adiabatic terms corresponding to the additional single-photon drive term introduced in \cref{eq:H-eps-w-phieps}.
For a weak single-photon drive, $\varepsilon(t)/\Omega(t) \ll 1$, the coherent states defining the eigenstates $\ket{C_\pm(\alpha)}$ are shifted by the small complex correction
\begin{equation}\label{eq:delta-alpha-def}
    \delta\alpha = \frac{\varepsilon(t)e^{-i\phi_\varepsilon}}{2K\alpha(t)^2}.
\end{equation}
The eigenstates for the KPO, including the single-photon drive, are now approximated cat states $\ket*{\tilde{C}_\pm}$ built from $\ket{\tilde{\alpha}_\pm} = \ket{\pm\alpha(t) + \delta\alpha}$. Working to the first order in $\delta\alpha$, an additional counteradiabatic terms appear beyond \cref{eq:H-counter-Nakamura} 
\begin{equation}\label{eq:H-counter-single-drive-phase}
\begin{split}
    \hat{H}_\mathrm{STA} &= \frac{i}{2} \frac{\varepsilon(t)\dot{\Omega}(t)}{\Omega(t)^2}\tanh\left(\frac{\Omega(t)}{K}\right)\left(e^{-i\phi_\varepsilon}\hat{a} - e^{i\phi_\varepsilon}\hat{a}^\dagger\right).
\end{split}
\end{equation}
In all our simulations, we consider the counteradiabatic terms \cref{eq:H-counter-Nakamura} and \cref{eq:H-counter-single-drive-phase} in addition to the primary generation Hamiltonian mentioned before; see \cref{app:sec:STA} for the more detailed derivation.

It also needs to mention that the STA calculation produces an extra term proportional to
$\varepsilon(t)\dot{\varepsilon}(t)/\Omega(t)^3$, but this term diverges at the beginning of the pulse and for $t>0$ is negligible compared to other terms, thus we neglect it from the simulation.

It is important to note that, for the dissipation-based generation model, the STA term used in the simulations differs from the Hamiltonian-based generation by an additional $\pi/2$ phase shift in the quadrature. This is because the Kerr term in the no-jump Hamiltonian has one extra phase factor $i$. Therefore, the terms in \cref{eq:H-counter-Nakamura} become proportional to $\propto(\hat{a}^{\dagger2}+\hat{a}^2)$. See Ref.~\cite{khanahmadi2025environment} for more details of the calculation in the dissipation-based case.

\section{Numerical characterization of flying cat qubits}\label{results}

In this section, we characterize the emitted flying cat qubits generated by the protocols introduced in \cref{subsec:generation_protocols}. We first study how the two-photon pump envelope controls the dominant emitted temporal mode in \cref{subsec:shaping-results}, followed by an analysis of logical rotations in \cref{subsec:X-rot,subsec:arb-rot-results} and noise sensitivity in \cref{subsec:noise-results}. Together, these results demonstrate a programmable source of flying cat qubits, providing a key building block for bosonic quantum networking and distributed quantum computing.

\subsection{Shaping propagating wave packets}\label{subsec:shaping-results}
High-fidelity release and capture of propagating quantum states require a well-matched sender and receiver; therefore, precise control of the emitted temporal mode is essential. Here, we demonstrate that our protocol provides control over the emitted temporal mode and enables shaping of the output field with high fidelity only by tuning the profile of the two-photon pump $\Omega(t)$.

As examples, we consider two profiles: an $n_\Omega$-th order low-pass filter (LPF) amplitude \cite{PhysRevA.99.023838},  and the second option is a symmetric shape. 

The LPF order described by the recursive relation
\begin{equation}\label{eq:Omega-m-def}
    \Omega_m(t) = \int_0^t B e^{-B(t-\tau)}\Omega_{m-1}(\tau) d\tau,
\end{equation}
where $\Omega_\mathrm{0}(t) = A e^{-\kappa_\mathrm{ex} t}$. Here, $A$ controls the drive amplitude and $B$ is the LPF bandwidth. Unless otherwise stated, we use the LPF pump with order $n_\Omega=3\rightarrow\Omega_3 \equiv \Omega(t)$ in our simulation. This drive has a fast ramp-up and a decay-like ramp-down, which makes it well-suited for our generation protocol.

Furthermore, to test a more symmetric release profile, we also introduce a sech-shaped pump envelope. The symmetric pulse is defined as
\begin{equation}\label{eq:Omega-sech-def}
    \Omega_\mathrm{sech}(t) = \frac{A}{\cosh{(\frac{t-t_c}{\sigma})}},
\end{equation}
with center $t_c=T_\mathrm{eff}/2$, and $T_\mathrm{eff}$ denotes the duration of the corresponding LPF pulse. 
\begin{figure}[h!]
    \centering
    \includegraphics[width=.93\linewidth]{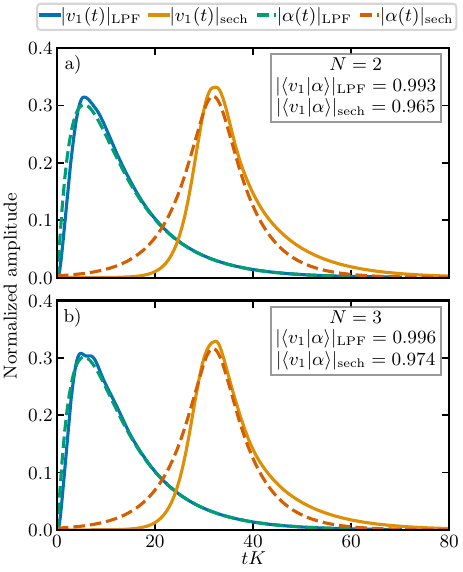}
    \caption{Dominant emitted mode $|v_1(t)|$ compared with the normalized cat amplitude profile $|\alpha(t)|$, where $\alpha(t)=\sqrt{\Omega(t)/K}$, for the Hamiltonian-based generation model. The rows correspond to $N=2$ and $N=3$. Curves are $L^2$-normalized, and labels report the overlap $|\braket{v_1}{\alpha}|$.} 
    \label{fig:mode-shaping}
\end{figure}

As mentioned, the amplitude of the cat state is proportional to the drive amplitude, $|\alpha|^2 \propto \Omega(t)$. The normalised $\alpha(t)$ and its corresponding $v_1$ mode of the output field are shown in \cref{fig:mode-shaping}, where the overlap $|\langle v_1|\alpha\rangle|$ determines the fidelity of the shaping protocol, reaching above $96\%$. The quantum state of each mode is also generated with high fidelity. For the LPF pulse, we obtain $\mathcal{F}=0.994$ for $N=2$ and $\mathcal{F}=0.993$ for $N=3$, while the sech-derived pulse gives $\mathcal{F}=0.989$ and $\mathcal{F}=0.988$, respectively. These fidelities are calculated according to the best cat-like state $\ket{\psi'}$ that maximize the fidelity $\mathcal{F}=\bra{\psi'}\hat \rho_{v_1}\ket{\psi'}$.

These results demonstrate highly efficient shaping control by only controlling the two-photon pump. In the remaining part of the paper, we consider the LPF pulse, as it has a faster ramp-up at the beginning, making it more relevant for simultaneous generation and release. It is also worth mentioning that achieving higher fidelity for the symmetric shape would require further optimization of the drive parameters, which we neglect here since it is not the focus of this work. 

\begin{figure}[t!]
    \centering
    \includegraphics[width=\linewidth]{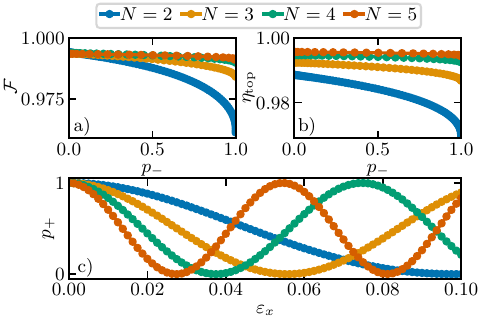}
    \caption{Logical $X$ rotation in the Hamiltonian-based generation model for $N=2,3,4,5$. (a) Cat-qubit fidelity $\mathcal{F}$ versus odd weight $p_-$. (b) Ratio of population  $\eta_\mathrm{top}=n_{v_1}/n_{\text{out}}$ versus $p_-$. (c) Even weight $p_+$ versus scaling factor $\varepsilon_x$. For more details see \cref{subsec:X-rot}.}
    \label{fig:LPF-orders-optimized}
\end{figure}
\subsection{Logical X-rotation on flying cat state}\label{subsec:X-rot}
While high-fidelity generation of flying cat states is important, the ability to control and manipulate the encoded logical state is equally essential. In the cat-state encoding, this corresponds to logical operations within the cat manifold, enabling control of both the logical basis populations and their relative phase through (X)- and (Z)-axis rotations, respectively.

To implement logical $X$-axis rotations, we use \cref{eq:H-eps-w-phieps} with $\phi_\epsilon=0$. The logical rotation angle is controlled by the amplitude of the single-photon drive. To conveniently parameterize its strength relative to the two-photon drive, we introduce the scaling factor $\varepsilon_x$. For low-pass-filter (LPF) orders ${n_\Omega,n_\varepsilon}$ corresponding to the two- and single-photon drives, respectively, the single-photon-drive envelope is defined as
\begin{equation}\label{eq:eps-def}
\varepsilon(t)=\varepsilon_x
\sqrt{\frac{\int_0^T |\Omega(t)|^2,dt}
{\int_0^T |\varepsilon(t;n_\varepsilon)|^2,dt}}
\,\varepsilon(t;n_\varepsilon),
\end{equation}
where $\varepsilon_x$ denotes the normalized amplitude of the single-photon drive. Unless otherwise stated, we use an LPF order of $n_\varepsilon=2$ for the single-photon-drive envelope.

\begin{figure*}[t!]
    \includegraphics[width=\linewidth]{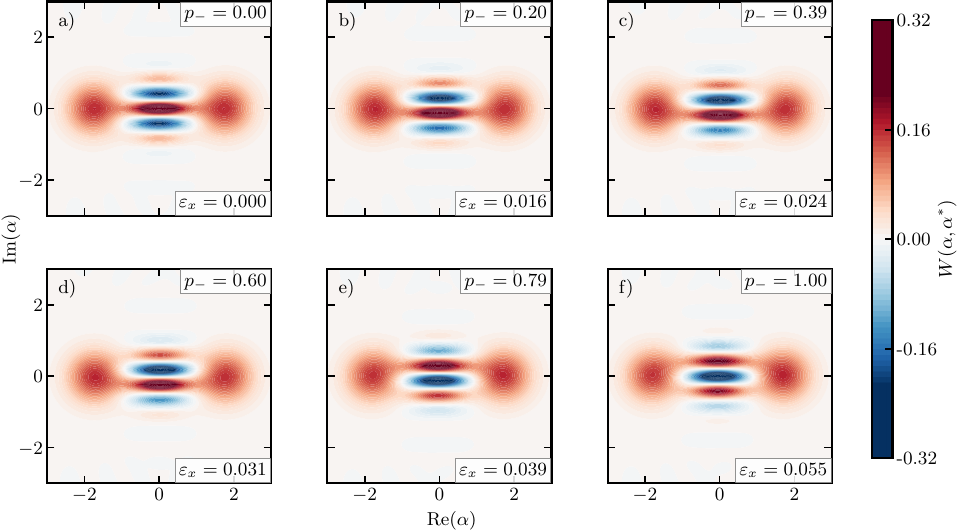}
    \caption{Wigner functions of the captured mode state $\hat{\rho}_{v_1}(T)$ with photon number $N=3$ shown in \cref{fig:LPF-orders-optimized}. The panels are selected along the $\varepsilon_x$ changing the quantum state from even-cat state ($p_-\approx 0.0$) to odd-cat state ($p_-\approx 1.0$). Each state is phase-rotated so that the cat axis is aligned with the $\mathrm{Re}(\alpha)$ axis.}
    \label{fig:Wigner-grid-Nout3}
\end{figure*}

We vary the amplitude of the single-photon drive within the range $0 \leq \varepsilon_x \leq 0.1$ to induce rotations from the even-cat state to the odd-cat state, as shown in \cref{fig:LPF-orders-optimized}, for output photon numbers $2 < N < 5$. Panels (a) and (b) show the fidelity and the fraction of the output photon population in $v_1$, $\eta_{\mathrm{top}} = n_{v_1}/n_{\mathrm{out}}$, respectively, as functions of the odd-cat-state population $p_-$. Throughout the $X$-rotation, the protocol maintains both high fidelity and an almost single-mode output field.

Panel (c) shows the resulting rotations as a function of the single-photon drive strength $\varepsilon_x$. Since the logical rotation angle is proportional to cat size $\propto\varepsilon_x |\alpha|$, states with larger photon numbers undergo multiple full rotations for the same drive ratio $\varepsilon_x$.

To visualize the logical rotation in phase space, \cref{fig:Wigner-grid-Nout3} shows Wigner functions generated in \cref{fig:LPF-orders-optimized} correspond to the cat state with $N=3$ photon number. The panels are selected along the first full rotation from $p+$ to $p_-$ according to increasing $\varepsilon_x$.

\begin{figure}[h!]
    \centering
    \includegraphics[width=\linewidth]{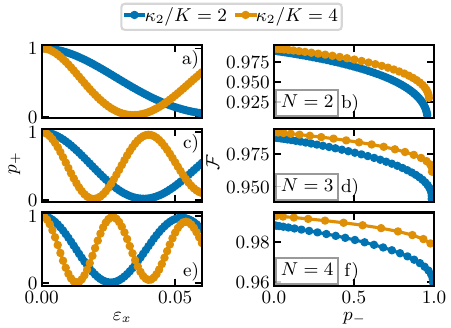}
    \caption{Logical $X$ rotation in the dissipation-based generation model for $N=2,3,4$ (rows). Different curves show different dissipative generation strengths $\kappa_2/K$. Left panels: Even weight $p_+$ versus $\varepsilon_x$. Right panels: Cat-qubit fidelity $\mathcal{F}$ versus odd weight $p_-$.}
    \label{fig:two-photon-loss}
\end{figure}

Furthermore, we also extend the X-rotation to the dissipation-based generation protocol for photon number $2<N<4$ with different two-photon dissipation rates as shown in \cref{fig:two-photon-loss}. We compare different dissipation rates $\kappa_2$ according to the Kerr nonlinearity $K$; the stronger dissipation $\kappa_2$, makes higher fidelity as the energy gap with the first excited state becomes larger and generation becomes more accurate. 

Finally, we introduce the optimization parameter $\lambda_\varepsilon$ to adjust the duration of the single-photon drive according to $\varepsilon(t)\rightarrow\varepsilon(\lambda_\varepsilon t)$. The optimization procedure and results are presented in \cref{app-subsec:lambda-eps}. We find that shorter single-photon pulses, relative to the two-photon drive, generally lead to higher-fidelity rotations. However, there is a trade-off between the drive amplitude and pulse duration which an optimal parameter regime can be identified numerically.

\subsection{Arbitrary rotation}\label{subsec:arb-rot-results}

\begin{figure}[h!]
    \centering    \includegraphics[width=\linewidth]{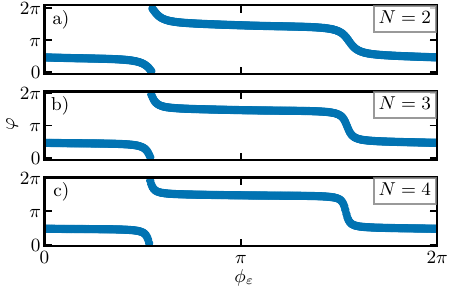}
    \caption{Cat-qubit relative phase control $\varphi$ with change of single-photon-drive phase $\phi_\varepsilon$. The relative phase $\varphi$ is shown versus the applied drive phase $\phi_\varepsilon$ for $N=2,3,4$ (rows) with the scaling factors $\varepsilon_x=4.5\%,\,2.75\%,\,1.75\%$, respectively, which give approximately balanced even/odd superpositions at $\phi_\varepsilon=0$. 
    }
    \label{fig:phi-vs-phieps}
\end{figure}

To have full control of the cat state, we now study the effect of changing the relative phase $\varphi$. This is achieved by varying the phase $\phi_\varepsilon$ of the single-photon drive in \cref{eq:H-eps-w-phieps}. For fixed scaling factors $\varepsilon_x$, we sweep the single-photon-drive phase $\phi_\varepsilon$ and extract the fitted relative phase $\varphi$ of the generated cat state, as shown in \cref{fig:phi-vs-phieps}, for photon number $N=2,3,4$. The initial state at $\phi_\varepsilon=0$ is the balanced even/odd superposition. 

However, we emphasize that varying the phase of the single-photon drive, $\phi_\varepsilon$, controls the superposition in the even- and odd-cat basis. To illustrate this, we show the resulting Bloch-sphere rotations for different drive amplitudes $\varepsilon_x$ and phases $\phi_\varepsilon$ in \cref{fig:app:phieps_epsx_scan,fig:app:phieps-epsx-scan-even-weight}, with additional discussion provided in \cref{app:Z-rot}.

\subsection{The effect of noises on encoding efficiency}\label{subsec:noise-results}

Any quantum protocol is subject to unwanted noise processes, most notably photon loss and pure dephasing, which can be modeled by the additional Lindblad operators
$\hat{L}_1=\sqrt{\kappa_1}\hat{a}$ and
$\hat{L}_\phi=\sqrt{\kappa_\phi}\hat{a}^\dagger\hat{a}$,
respectively. We consider noise strengths up to $1\%$ of the dissipation rate $\kappa_\mathrm{ex}$. For waveguide coupling rates of a few MHz, this corresponds to decoherence rates on the order of several tens of kHz, which are relevant to current experimental superconducting-circuit platforms \cite{reglade2024quantum}.
Additionally, we introduce an extra optimization variable $\gamma_\Omega$ which scales the width of the two-photon pump as $\Omega(t)\rightarrow\Omega(\lambda_\Omega t)$, so that $\lambda_\Omega>1$ corresponds to a temporally compressed pump and $\lambda_\Omega<1$ to a stretched pump. 

\begin{figure}[h!]
    \centering
    \includegraphics[width=\linewidth]{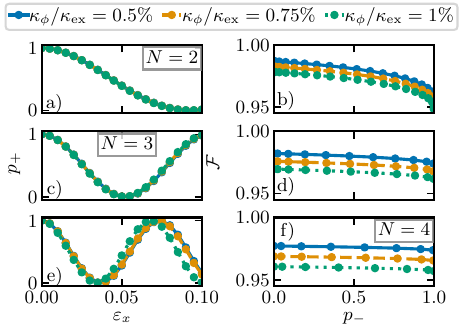}
    \caption{Dephasing effects on cat control process, in the Hamiltonian-based generation model for $N=2,3,4$ (rows). Left: Even weight $p_+$ versus $\varepsilon_x$. Right: Cat-qubit fidelity $\mathcal{F}$ versus $p_-$. Colors indicate $\kappa_\phi/\kappa_\mathrm{ex}=0.5\%,\,0.75\%,\,1\%$. }
    \label{fig:dephasing-Kerr-model}
\end{figure}

\begin{figure}[h!]
    \centering
    \includegraphics[width=\linewidth]{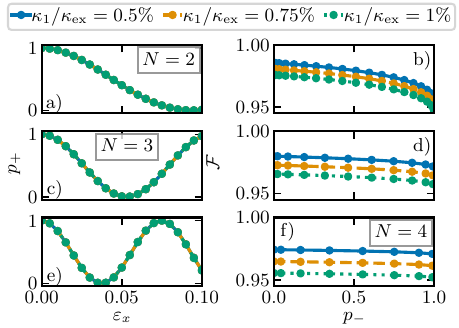}
    \caption{Single-photon-loss effects on cat control process, in the Hamiltonian-based generation model for $N=2,3,4$ (rows). Left: Even weight $p_+$ versus $\varepsilon_x$. Right: Cat-qubit fidelity $\mathcal{F}$ versus $p_-$. Colors indicate $\kappa_1/\kappa_\mathrm{ex}=0.5\%,\,0.75\%,\,1\%$. The optimized two-photon pump scaling was $\lambda_\Omega=1.1$ for $N=2,3$ and $\lambda_\Omega=1.0$ for $N=4$.}
    \label{fig:onephoton-loss-Kerr-model}
\end{figure}

The dephasing and photon loss on Hamiltonian-based generation are shown in \cref{fig:dephasing-Kerr-model} and \cref{fig:onephoton-loss-Kerr-model}, respectively. The results show fidelities above $95\%$ throughout the rotation gate for photon number $N=\{2,3,4\}$.  However, the error has a stronger effect at higher photon numbers, affecting both the modeness of the output field and the corresponding fidelity. To increase the fidelity, we optimized the parametric-drive parameter $\lambda_\Omega$ and found the optimal value $\lambda_\Omega=1.1$ for \cref{fig:dephasing-Kerr-model}, and for \cref{fig:onephoton-loss-Kerr-model}, $\lambda_\Omega=\{1.1$ and $1.0\}$ for $N=\{\{2,3\}$ and $4\}$, respectively.

\begin{figure}[h!]
    \centering
    \includegraphics[width=\linewidth]{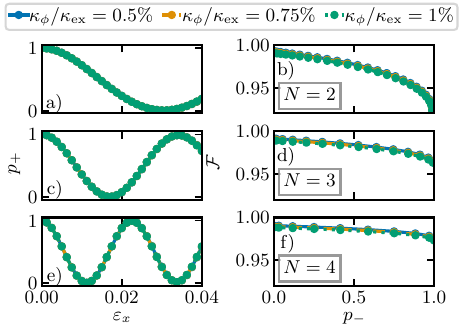}
    \caption{Dephasing effects on cat control process in the dissipation-based generation model with $\kappa_2/K=4$ for $N=2,3,4$ (columns). Left panels: Even weight $p_+$ versus $\varepsilon_x$. Right panels: cat-qubit fidelity $\mathcal{F}$ versus $p_-$. Marker shape indicates $\kappa_\phi/\kappa_\mathrm{ex}=0.5\%,\,0.75\%,\,1\%$. The optimized pump scaling was $\lambda_\Omega=1.3$ in all cases.}
    \label{fig:dephasing-NOKerr-model}
\end{figure}

\begin{figure}[h!]
    \centering
    \includegraphics[width=\linewidth]{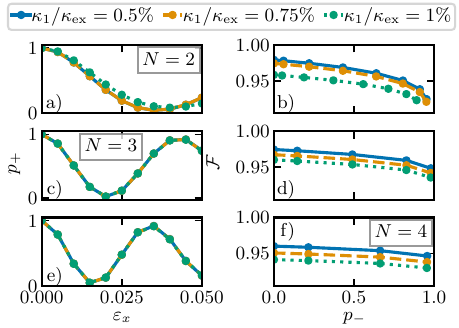}
    \caption{Single-photon-loss effects on cat control process, in the dissipation-based generation model with $\kappa_2/K=4$ for $N=2,3,4$ (rows). Left: Even weight $p_+$ versus $\varepsilon_x$. Right: Cat-qubit fidelity $\mathcal{F}$ versus $p_-$. Colors indicate $\kappa_1/\kappa_\mathrm{ex}=0.5\%,\,0.75\%,\,1\%$. Single-photon loss limits the achievable rotation, preventing complete transfer from the even- to the odd-cat state at larger $\varepsilon_x$; see \cref{subsec:noise-results} for more details.}
    \label{fig:onephoton-loss-NOKerr-model}
\end{figure}

The results of noise on dissipation-based generation protocol are illustrated in \cref{fig:dephasing-NOKerr-model} and \cref{fig:onephoton-loss-NOKerr-model}.
Here, the dissipative generation strength is fixed at $\kappa_2/K=4$. The optimized parameter is found to be $\lambda_\Omega=1.3$ for dephasing noise, while for single-photon loss the optimal values are $\lambda_\Omega=0.9$ for $N=\{2,3\}$, and $\lambda_\Omega=0.7$ for $N=4$.

Thus, it should be noted that, in the presence of single-photon loss, we find that a slower drive ($\lambda_\Omega<1$) can be optimal for generating high-fidelity cat states. This observation highlights the importance of balancing noise-induced phase changes, the single-mode character of the emitted state, and the fidelity of the generated state. 
Furthermore, we note that single-photon loss has a stronger impact on the full $X$-rotation for larger $\varepsilon_x$, as shown in \cref{fig:onephoton-loss-Kerr-model,fig:onephoton-loss-NOKerr-model}. Single-photon loss changes the parity of the cat state and therefore affects later stages of the rotation dynamics, particularly when multiple rotation cycles are involved. In contrast, the first rotation from the even-cat state to the odd-cat state is only weakly affected, and the dynamics remain mostly unchanged.

Furthermore, we note that, in all fidelity-versus-$p_-$ results, the cat state with $N=2$ shows a larger reduction in fidelity near the complete rotation to the odd-cat state than states with higher photon numbers. This arises from the fact that the logical rotation angle scales as $\propto\varepsilon(t)|\alpha|$. Consequently, for smaller cat amplitudes, a stronger single-photon drive is required to achieve the same rotation. As the drive strength increases, the weak-drive approximation, i.e. $\varepsilon(t)\ll\Omega(t)$, becomes less accurate, leading to leakage outside the logical cat subspace and a corresponding reduction in fidelity. This effect is significantly suppressed for larger cat states, where the required drive amplitude is smaller.

\section{Summary}\label{summary}

We have presented a protocol for the programmable generation and logical control of flying two-component cat qubits emitted into a waveguide. In our protocol, the source oscillator remains continuously coupled to the waveguide, allowing cat-state generation, emission, and logical manipulation to occur simultaneously. We investigated both Kerr-based and two-photon-dissipation-based generation schemes, both of which are compatible with superconducting-circuit platforms.

Our results show that high-fidelity temporal-mode shaping of the emitted cat qubit can be achieved using only a time-dependent two-photon pump. 
In addition, by applying a single-photon drive with an optimized amplitude, temporal profile, and phase, we obtain full control over logical rotations of the emitted bosonic qubit. This demonstrates that logical operations can be performed while the quantum information is shared between the source oscillator and the propagating field, rather than requiring the complete state to be localized in the oscillator. Such hybrid stationary--flying control opens the possibility of performing in-flight processing of bosonic qubits, which could be advantageous for modular quantum networks, quantum repeater architectures, and distributed bosonic quantum computing.

We also find that the protocol remains functional with high fidelity under weak pure dephasing and additional single-photon loss. Optimizing the duration of the two-photon pump partially mitigates the effect of noise and improves the robustness of the emitted cat-qubit encoding. The robustness of the protocol may be further enhanced by combining Kerr nonlinearity with engineered dissipation. Such combined Kerr--dissipation schemes have been shown to improve the robustness of stationary cat-state generation \cite{Gautier_2022} and may provide similar advantages for the generation and emission of propagating cat qubits.

Future work could extend this approach in several directions. A first step is to generalize the protocol to error-correctable bosonic encodings, such as four-component cat codes \cite{joshi2021quantum} and pair-cat states \cite{albert2019pair,PhysRevA.106.062422,PRXQuantum.4.020319}. Since these states require higher-order nonlinear processes, the present protocol will likely need to be combined with time-dependent optimal-control methods. More broadly, the present framework may be extended to realize universal logical control of propagating bosonic qubits, providing a route toward more programmable and robust bosonic quantum links between superconducting-circuit nodes.

\section{Acknowledgment}
This work was supported by the Knut and Alice Wallenberg Foundation through the Wallenberg Centre for Quantum Technology (WACQT). C. E. and M. K. acknowledge the support from the Marianne and Marcus Wallenberg Foundation (Grant No. MMW 022.0167) through the EDU-WACQT Summer Internship Program. Z. Z. and G. J. acknowledge support from the Swedish Research Council (Grant No. 2021-04037). The simulations were enabled by resources provided by the National Academic Infrastructure for Supercomputing in Sweden (NAISS), partially funded by the Swedish Research Council through grant no. 2022-06725.


\appendix

\section{Shortcut to adiabaticity Hamiltonian}\label{app:sec:STA}

We derive the counteradiabatic Hamiltonian from \cref{eq:H-counter-Nakamura,eq:H-counter-single-drive-phase}. The starting point is the KPO Hamiltonian with the single-photon drive
\begin{equation}\label{eq:app-H-KPO}
\begin{split}
    \hat{H}_\mathrm{KPO} = \frac{\Omega(t)}{2}\left(\hat{a}^{\dagger 2} + \hat{a}^2\right) &- \frac{K}{2}\hat{a}^{\dagger 2}\hat{a}^2 \\ &+ \varepsilon(t)\left(e^{-i\phi_\varepsilon}\hat{a}^\dagger + e^{i\phi_\varepsilon}\hat{a}\right),
\end{split}
\end{equation}
We will assume a weak single-photon drive such that $\varepsilon(t)/\Omega(t) \ll 1$ and a constant single-photon drive phase $\phi_\varepsilon$.

\subsection{Approximate cat states}
For $\varepsilon(t) = 0$, the eigenstates of $\hat{H}_\mathrm{KPO}$ from \cref{eq:app-H-KPO} are even and odd cat states $\ket{C_\pm(\alpha)}$ built from the coherent states $\ket{\pm\alpha}$ with $\alpha = \sqrt{\Omega(t)/K}$. In the presence of a weak single-photon drive, we assume that the eigenstates of $\hat{H}_\mathrm{KPO}$ are built from approximate coherent states on the form $\ket{\tilde{\alpha}_\pm(t)} = \ket{\pm\alpha(t) + \delta\alpha(t)}$. By applying the variational method, i.e. solving $\frac{d}{d\tilde{\alpha}} \expval{H_\text{KPO}}{\tilde{\alpha}} = 0$, we find the small complex correction
\begin{equation}\label{eq:app-beta-pm}
    \delta\alpha(t) = \frac{\varepsilon(t)e^{-i\phi_\varepsilon}}{2K\alpha(t)^2},
\end{equation}
and the approximate eigenstates to first order in $\delta\alpha(t)$ are
\begin{equation}\label{eq:app-cat-states}
    \ket*{\tilde{C}_\pm(t)} = \frac{\ket{\tilde{\alpha}_+(t)} \pm \ket{\tilde{\alpha}_-(t)}}{\mathcal{N}_\pm},
\end{equation}
with normalization constant $\mathcal{N}_\pm = \sqrt{2\left(1 \pm e^{-2\Omega(t)/K}\right)}$.

\subsection*{Time derivative of the approximate eigenstates}

To evaluate the STA Hamiltonian of \cref{eq:H-STA}, we begin by evaluating the time derivative of the approximate eigenstates
\begin{equation}
    \ket*{\dot{\tilde{C}}_\pm} = \frac{1}{\mathcal{N}_\pm}\frac{d}{dt}\left(\ket*{\tilde{\alpha}_+} \pm \ket*{\tilde{\alpha}_-}\right) - \frac{\dot{\mathcal{N}}_\pm}{\mathcal{N}_\pm}\ket*{\tilde{C}_\pm},
\end{equation}
where
\begin{equation}\label{eq:app-Ndot}
    \dot{\mathcal{N}}_\pm = \mp\frac{2\dot{\Omega}(t)e^{-2\Omega(t)/K}}{K\sqrt{2\left(1 \pm e^{-2\Omega(t)/K}\right)}}.
\end{equation}
For the derivative of a coherent state with amplitude $\beta$ we use the identity $\tfrac{d}{dt}\ket{\beta(t)} = (\dot\beta\hat{a}^\dagger - \dot\beta^*\hat{a})\ket{\beta(t)}$. Furthermore, we take use of
\begin{equation}\label{eq:app-Nratio}
        \frac{\mathcal{N}_\mp}{\mathcal{N}_\pm} = \sqrt{\tanh^{\pm 1}\left(\frac{\Omega(t)}{K}\right)},
\end{equation}
and
\begin{equation}\label{eq:app-a-act}
    \hat{a} \left(\ket{\tilde{\alpha}_+} \pm \ket{\tilde{\alpha}_-}\right) = \alpha(\ket{\tilde{\alpha}_+} \mp \ket{\tilde{\alpha}_-}) + \delta\alpha(\ket{\tilde{\alpha}_+} \pm \ket{\tilde{\alpha}_-}),   
\end{equation}
to find that
\begin{equation}\label{eq:app-Cdot-form}
    \ket*{\dot{\tilde{C}}_\pm} = A \ket*{\tilde{C}_\pm} + B \hat{a}^\dag \ket*{\tilde{C}_\mp} +  C \hat{a}^\dag \ket*{\tilde{C}_\pm} + D \ket*{\tilde{C}_\mp},
\end{equation}
with coefficients
\begin{align}
    A &= -\frac{\dot{\Omega}(t)}{2K}\tanh^{\pm 1}\left(\frac{\Omega(t)}{K}\right) - \dot{\delta\alpha}^*\delta\alpha, \label{eq:app-A} \\
    B &= \dot\alpha\sqrt{\tanh^{\pm 1}\left(\frac{\Omega(t)}{K}\right)}, \label{eq:app-B} \\
    C &= \dot{\delta\alpha}, \label{eq:app-C} \\
    D &= -\left(\dot\alpha\delta\alpha + \dot{\delta\alpha}^*\alpha\right)\sqrt{\tanh^{\pm 1}\left(\frac{\Omega(t)}{K}\right)}. \label{eq:app-D}
\end{align}

\subsection{STA Hamiltonian}
The contributions from $\ket*{\tilde{C}_-}$ in \cref{eq:H-STA} are discarded as they diverge at $t = 0$. Retaining only the positive cat terms gives us that,
\begin{equation}
\begin{split}
    \hat{H}_\mathrm{STA} &= \frac{i}{2}\left(\ket*{\dot{\tilde{C}}_+}\bra*{\tilde{C}_+} - \ket*{\tilde{C}_+}\bra*{\dot{\tilde{C}}_+}\right) \\
    &= -\frac{i}{2}\ket*{\tilde{C}_+}\bra*{\dot{\tilde{C}}_+} + \mathrm{h.c.}
\end{split}
\end{equation}
Now we will evaluate $\hat{H}_\mathrm{STA}$ acting on $\ket*{\tilde{C}_+}$ by using that
\begin{equation}
\begin{split}
    \bra*{\tilde{C}_-}\hat{a}\ket*{\tilde{C}_+} &= \alpha \sqrt{\tanh\left(\tfrac{\Omega(t)}{K}\right)}, \\
    \bra*{\tilde{C}_+}\hat{a}\ket*{\tilde{C}_+} &= \delta\alpha.
\end{split}
\end{equation}
one finds that
\begin{equation}\label{eq:app-Hcounter-onC+}
    \hat{H}_\text{STA} \ket*{\tilde{C}_+} = -i\Big(\dot{\alpha} \alpha\tanh\left(\tfrac{\Omega(t)}{K}\right) + \dot{\delta\alpha}^*\delta\alpha \Big) \ket*{\tilde{C}_+} + \mathrm{h.c.}
\end{equation}
Finally by using the eigenvalue relation
\begin{equation}
    \ket*{\tilde{C}_\pm} = \frac{\hat{a}^2 - 2\delta\alpha \hat{a}}{\alpha^2} \ket*{\tilde{C}_\pm} + \mathcal{O}(\delta\alpha^2),
\end{equation}
we find that
\begin{equation}
\begin{split}
    \hat{H}_\text{STA} = \frac{i}{2}\bigg[&\frac{\dot{\Omega}(t)}{\Omega(t)}\tanh\left(\frac{\Omega(t)}{K}\right)\left(\hat{a}^{\dagger 2}-\hat{a}^2\right) \\
    &+ \frac{\varepsilon(t)\dot{\varepsilon}(t)K}{2\Omega(t)^3}\left(\hat{a}^{\dagger 2}-\hat{a}^2\right) \\ &+ \frac{\varepsilon(t)\dot{\Omega}(t)}{\Omega(t)^2}\tanh\left(\frac{\Omega(t)}{K}\right)\left(e^{-i\phi_\varepsilon}\hat{a} - e^{i\phi_\varepsilon}\hat{a}^\dagger\right)\bigg]
\end{split}
\end{equation}

The second term in this expression diverges at the beginning of the protocol, where $\Omega(t)\rightarrow0$. The second term is therefore omitted, which gives the counteradiabatic Hamiltonians in \cref{eq:H-counter-Nakamura,eq:H-counter-single-drive-phase}.

\section{Baseline simulation parameters}\label{app:baseline-parameters}
This section summarizes the baseline numerical parameters used in the simulations. All quantities are reported in dimensionless units with $K=1$, where $K$ denotes the reference frequency scale. Throughout, the external waveguide coupling is fixed to $\kappa_\mathrm{ex}=0.2K$, and the LPF bandwidth is chosen as $B=2.5\kappa_\mathrm{ex}$. The total simulation time is $T=70/K$, so that the dimensionless simulation interval is $Kt\in[0,70]$. This time window is used both for computing the output-field correlation function and for the subsequent virtual-cavity capture. 

\begin{figure}[h!]
    \centering
    \includegraphics[width=0.85\linewidth]{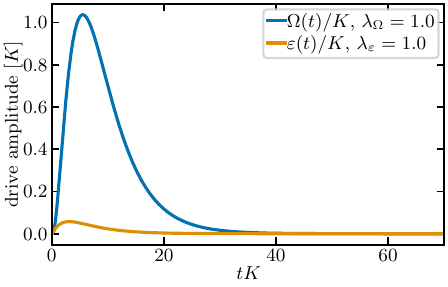}
    \caption{Example two-photon pump envelope $\Omega(t)$ and single-photon drive envelope $\varepsilon(t)$ for $\lambda_\Omega=\lambda_\varepsilon=1$, used to generate a cat state with photon number $N=2$ and $\varepsilon_x=0.05$.}
    \label{fig:drive-widths-p-eps}
\end{figure}

We show one example of the temporal profile of the two- and single-photon drive in, \cref{fig:drive-widths-p-eps}, with parameter $\lambda_\Omega=\lambda_\varepsilon=1$, using LPF orders $(n_\Omega,n_\varepsilon)=(3,2)$, $N=2$ and $\varepsilon_x=0.05$. As mentioned in the main text, the initial state of the source oscillator is the vacuum state. For simulations targeting a given output photon number $N$, the pump amplitude $A$ is determined numerically such that the final photon number captured in the virtual cavity satisfies $N_\mathrm{cap}(T)\approx N$.

\begin{figure}[h!]
    \centering
    \includegraphics[width=\linewidth]{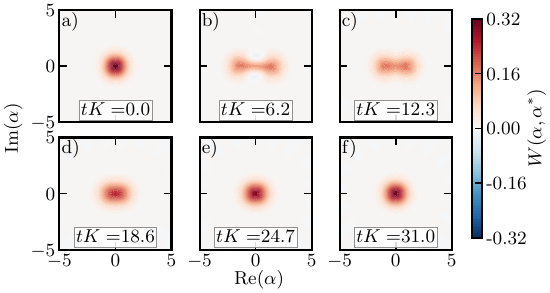}
  \caption{Wigner functions of the source-oscillator state $\rho_\mathrm{s}(t)$ at selected times during the generation of a cat state with photon number $N=2$ with the drive shown in \cref{fig:drive-widths-p-eps}. Although the emitted wavepacket approaches a high-fidelity cat state, the intracavity state never evolves into a well-defined cat state because generation and emission occur simultaneously.}
    \label{fig:app:intracavity-dynamics}
\end{figure}

\section{Intracavity dynamics during generation and release} \label{app:intracavity-dynamics}
This section provides additional information and characteristics of the intra-cavity quantum state evolution during the generation of the travelling cat state. 

To visualize the intracavity dynamics, we plot the Wigner function of the source-oscillator state $\hat{\rho}_\mathrm{s}(t)$ at selected times during the Kerr-based generation protocol in \cref{fig:app:intracavity-dynamics}. According to the drive profiles shown in \cref{fig:drive-widths-p-eps}, the drives gradually transform the vacuum state into a small cat-like state during the generation process. However, because the quantum state is continuously emitted into the waveguide, the intracavity state never evolves into a well-defined cat state at any time.

\section{Logical $Z$ rotation}\label{app:Z-rot}

\begin{figure}
    \centering
    \includegraphics[width=\linewidth]{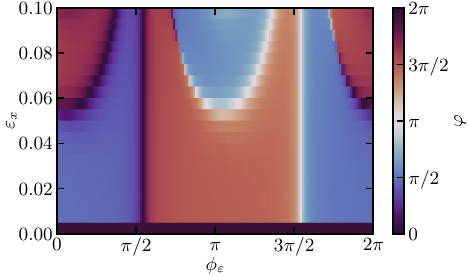}
    \caption{Cat-qubit phase control over a two-dimensional sweep of single-photon-drive amplitude and phase. The color scale shows the relative phase $\varphi$ as a function of $\varepsilon_x$ and $\phi_\varepsilon$ for fixed pump amplitude corresponding approximately to photon number $N=3$.}
    \label{fig:app:phieps_epsx_scan}
\end{figure}

\begin{figure}
    \centering
    \includegraphics[width=\linewidth]{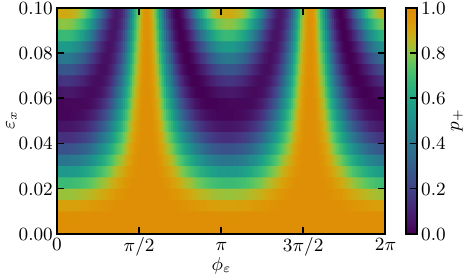}
  \caption{Even-cat population $p_+$ as a function of the single-photon-drive amplitude $\varepsilon_x$ and phase $\phi_\varepsilon$. The color scale illustrates the control of the logical-state population over the same parameter sweep shown in \cref{fig:app:phieps_epsx_scan}. }
    \label{fig:app:phieps-epsx-scan-even-weight}
\end{figure}

For a fixed two-photon pump amplitude chosen to give approximately $N=3$, we scan the scaling factor $\varepsilon_x$ and drive phase $\phi_\varepsilon$. The resulting logical phase $\varphi$ and even population $p_+$ are shown in \cref{fig:app:phieps-epsx-scan-even-weight,fig:app:phieps_epsx_scan}, respectively. Varying $\phi_\varepsilon$ tunes the emitted cat-qubit phase over the full interval $[0,2\pi]$. For very small $\varepsilon_x$, the state remains close to the even-cat pole, so the relative phase is not physically well defined; once the odd component is populated, $\varphi$ becomes meaningful. By comparing these two figures \cref{fig:app:phieps-epsx-scan-even-weight,fig:app:phieps_epsx_scan}, one can determine the appropriate amplitude $\varepsilon_x$ and phase $\phi_\varepsilon$ required to generate the desired cat-state superposition within the logical subspace.


\begin{figure*}[ht!]
    \centering
    \includegraphics[width=\linewidth]{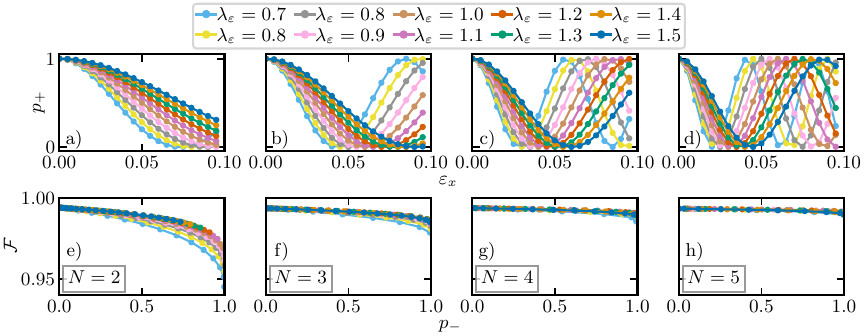}
    \caption{Optimization of the single-photon-drive time scaling factor $\lambda_\varepsilon$ for $N=2,3,4,$ and $5$ (columns). The two-photon pump scaling is fixed at $\lambda_\Omega=1$. Top: even-cat weight $p_+$ as a function of $\varepsilon_x$. Bottom: cat-qubit fidelity $\mathcal{F}$ as a function of the odd-cat weight $p_-$.}
    \label{fig:lambda-eps}
\end{figure*}

\section{Time-scaling optimization of the single-photon drive pulse}\label{app-subsec:lambda-eps}
As discussed in the main text, the profile of the single-photon drive is optimized through the scaling factor $\lambda_\varepsilon$. Here, we investigate the effect of varying $\lambda_\varepsilon$ on the fidelity and logical $X$-rotation shown in \cref{fig:lambda-eps} for photon numbers $N=2,3,4,$ and $5$.

For shorter pulses ($\lambda_\varepsilon>1$), a larger drive amplitude $\varepsilon_x$ is required to achieve a full logical rotation, whereas longer pulses require weaker drive amplitudes. Therefore, there is a trade-off between the pulse duration and the drive strength. Depending on the experimental constraints and noise environment, an appropriate scaling factor can be chosen to optimize the protocol performance.



\bibliographystyle{apsrev4-2}

\end{document}